# Preparing students for authentic research through open-inquiry laboratory courses

**Roxanne P. Spencer**[*], **Sergey V. Samsonau**

*STEAM Department, Princeton International School of Mathematics and Science, Princeton, NJ, USA*

**Abstract**

Authentic research as a method of teaching science is gaining popularity in high schools and colleges. To make this research experience most efficient, students need adequate preparation in traditional science courses. Existing materials available for inquiry-based teaching methods are not intended to prepare a student for independent research. We discuss principles for organization of laboratory classes with this goal in mind and provide examples that we have implemented for algebra-based physics, calculus-based physics, and honors chemistry for high school students.

## 1   INTRODUCTION

Science, Technology, Engineering, and Math (STEM) education will be essential for the US economy in the 21st century (President's Council of Advisors on Science and Technology, 2010). To ensure that the US remains a leader in the STEM field, many have focused attention on the need for improvements in STEM education in the United States. A recent report by the President's Council of Advisors on Science and Technology (PCAST) called for the support of state-led standards in STEM curricula, the recruitment of 100,000 new STEM teachers, and the creation of 1,000 new STEM-focused schools (President's Council of Advisors on Science and Technology, 2010). The growing concern over the state of STEM education in the United States is clearly warranted. According to a report published by Consortium for Student Retention Data Exchange, out of students who begin as a STEM major only 46.2% will graduate in 6 years (65.0% if we account for those who switched to a non-STEM major).

Two recent efforts to improve STEM education include developing inquiry-oriented curriculum and increasing access to authentic research programs for students. Inquiry-oriented education has been shown to provide numerous advantages over traditional instruction-based approach (Barron and Darling-Hammond, 2018; Gormally et al., 2009). Unlike lecture-based approaches, inquiry is intended to have students mimic the actual scientific process by "doing" science in a more authentic context (Fay and Bretz, 2008; Lederman et al., 2014; Moebius-Clune et al., 2011). This allows students to learn science as both a body of knowledge and a method of knowing. Authentic research programs, though similar in general approach to inquiry, are longer term (typically, from 1 to 3 years) programs that allow students to work on real scientific questions for a significant period of time. Such programs are being introduced across the country in various public and private high schools such as Thomas Jefferson High School of Science and Technology (Hannum, 2016), Laguna Beach High School (Sogo, 2016), and Princeton International School of Mathematics and Science (PRISMS), among others (National Research Council (U.S.), 2011). Various organizations have been established to coordinate the efforts, such as the National Consortium of Secondary Stem Schools (NCSSS) and the Global Sphere Network. This effort is based on studies showing significant benefits for students participating in such programs (National Research Council (U.S.), 2011; Rodenbusch et al., 2016).

---

[*] Corresponding author: Roxanne P. Spencer, rspencer@alumni.princeton.edu

Here we would like to discuss interconnecting these two movements so that classes relying on inquiry methods can prepare students for a subsequent authentic research experience. Further, the practices gained from such inquiry methods can enable students meet the key science and engineering practice goals articulated in the Next Generation Science Standards (National Research Council (U.S.), 2012). Even though the effectiveness of such preparation has yet to be assessed formally, our experience suggests that students who undergo such preparation perform better during their research.

In this paper we focus on the laboratory component of science classes. While many high school and colleges have introduced inquiry-oriented labs, we point out that the term "inquiry" is sometimes loosely used (Buck et al., 2008) and can be applied indiscriminately to activities ranging from limited flexibility in interpreting the data to obtain an answer to wholly-student designed procedures. It is up to the individual instructor to interpret "inquiry" and the level of "flexibility" to incorporate into labs (Backus, 2005; Fay and Bretz, 2008). On top of that, the variety of topics, instructor experience, and available equipment mean that there is no single set of labs that meet every pedagogical goal. Thus, an instructor aiming to introduce open-inquiry curriculum must review many examples and build classes "by analogy" to fit a particular situation. There are various sets of guiding principles described for various learning goals (Backus, 2005; Grady, 2010; Meyer and Avery, 2010; Rodríguez-Arteche and Martínez-Aznar, 2016; Walker et al., 2011).

To the best of our knowledge, however, no approach has been proposed for the design of inquiry labs intended to prepare students for authentic research; herein we present a set of guiding principles for such curriculum. We use the Harwood model of inquiry (Harwood, 2004b,a) to evaluate laboratory sequences for physics and chemistry to ensure that they provide opportunities for students to develop the necessary skills for meaningful scientific inquiry.

## 2 GUIDING PRINCIPLES FOR INDEPENDENT RESEARCHER LABS

### 2.1 Work of a student in a typical research lab

While focusing on preparing a student for research work, we first consider the typical experience of a student starting work in a research lab, namely, the probable reproduction of an existing protocol under the guidance of a lab director. Alternatively, a student may be asked to determine the limit of the applicability of a given model. These student tasks can follow many approaches; however, it is highly unlikely that a beginning student would be given a totally unknown effect to study or required to formulate a hypothesis from scratch (thus, fully "open ended labs" do not seem to be useful in teaching beginning students).

We can summarize the steps a beginning researcher might follow as (1) get a general or specific task from a senior researcher, (2) read and evaluate literature methods, (3) expecting that reproduction of existing procedures will be fast and simple, (4) obtain results, both positive outcomes and failures and discover that things often take longer than expected, (5) recognize and analyze reasons for those failures, (6) re-read and re-evaluate the methods from the literature; (7) adapt and revise an existing procedure; and repeat the process until the student decides that a result is sufficiently good.

### 2.2 Formulating a task

A central theme of research is that tasks are not always fully elucidated and are often vague. Experimentation is more than simply following a recipe and requires reasoning and analysis. For example, students can be asked to determine concentration of ingredients in aspirin, measure the resistivity of a wire, or determine efficiency of a solar cell; even though an approximate value may be given by a manufacturer, the specific value can vary slightly from lot to lot and is essentially unknown. Sometimes there is no information from a manufacturer at all.

Experiments can be based on a well-known experiment described in a textbook or discussed in class with at least one added complication, or "catch", that we expect students to recognize and address. This may be as simple as low precision instruments or other equipment limitations. Students need to learn to recognize when something went wrong or won't work, correct their experimental designs, and obtain a result that they can independently consider acceptable. Note that we emphasize reasonable results rather than correct results; the "answer" is often unknown in research.

A student presented with only a general formulation of a task must rely on subject knowledge to create a procedure. Thus, the methods may vary from group to group and influence the results such that there may be widely distinct results among groups. This allows students to learn that there is no "one right method" and that different methods may lead to different results in terms of precision or accuracy, for example.

## 2.3  Non-ideal experimentation

In some cases, the labs are presented using a "first draft approach" in which students develop an experimental protocol to achieve an assigned objective (for example, verification of a law or characterization of a material). In such draft experiments, there are many sources of errors that students have to identify and evaluate. As part of their experimental write-up, they describe how to improve the experimental design to achieve a better quality of the results.

Students discover that while a solid theoretical understanding is necessary for practical applications, a procedure that sounds good conceptually can fail in practice. For example, a common issue is an initial experimental design that requires precision not attainable with available equipment. Or, the inability to completely eliminate external factors from influencing a measurement. This is realistic preparation for working in the research lab: straightforward procedures often involve many nuances that require significant effort to master.

Students must use their judgment to decide whether they have obtained a reasonable and sufficient result. If they decide that the result is not good enough, they can redesign and improve their procedure. Students are given the opportunity to redo labs (for instance, during the next class or office hours) and are not penalized for experimental failures. Indeed, making mistakes and recognizing them—even after the fact—is considered one of the key aspects of inquiry-based education (Moebius-Clune et al., 2011). As a study on how students deal with unexpected data in the chemistry lab concludes that it is important for students to have opportunities to discuss limitations in data, such as systematic errors (Pickering and Monts, 1982).

## 2.4 Self-study and focus on educational environment

In most classes, students are conditioned to expect that new topics will be officially introduced and taught. Often in research, some knowledge is learned as needed, and researchers often must think on their feet. Thus, for some labs, students may have to apply knowledge that was not formally covered in class; students are given suggestions where to find information, but the depth and quantity of external literature research is left to the individual student; in general, students with deeper understanding produce deeper analysis and better experiments. This allows differentiation in the labs as students progress at their level and develop a sense of responsibility for their knowledge. Students are taught to explore additional sources of information (in essence, to read the literature) and see that there is always something else to learn.

The teacher must create an environment where students can explore nature, wonder, and experiment, as well as receive a feedback about successes and failures, rather than simply focusing on covering subject material for a test.

## 2.5 Continuous teacher innovation

We recommend teachers prepare at least one new lab every semester for which the teacher is deliberately not "overprepared" (by this, we mean that a teacher may not know all potential ways that an experiment may fail; in all cases, the teacher has completed a safety analysis of any proposed student experimental work). This requires the teacher to innovate and maintain subject knowledge, skills and interest. It also reduces the student institutional knowledge in which students seek out lab reports from previous years, rather than confront the material on their own.

By doing such a lab with a class, a teacher will encounter unexpected problems and must decide how to overcome those in front of students, thus providing an illustration of the need to adapt. We think that it is important for students to see how a teacher deals with failures to model for students how to deal with frustration. Additionally, students see there is room for error and uncertainty and that no one has all the answers. We want to create an atmosphere in which questioning a more experienced scientist or teacher is acceptable.

**2.6 Lab reports and peer review**

Instead of using an arbitrary lab report format, our students write lab articles following the style of a typical research journal article with abstract, introduction, experimental materials and methods section, results, discussion, and conclusion. Students are expected to include correctly formatted references. For consistency across disciplines in our school,

We require our students use a single citation style based on the *Science* Citation Style (American Association for the Advancement of Science, 2018). We encourage our students to use a reference manager (we typically use Mendeley; there are several programs available such as Zotero and Endnote, among others), and offer a citation style language (CSL) file with the format. The emphasis on standard writing formats and styles is important to train students how to communicate results and improves their understanding of scientific papers.

Access to the scientific literature is essential in our model. While many students ask for an example to follow, this was found to be ineffective since students tend to simply replace content with their own, and not think how to organize material. We ask students to find examples of regular scientific papers that are relevant to the subject area to learn the general principles that scientists use to write papers rather than focusing on a specific teacher's preferences. There are various free and subscription databases that allow students to access papers (Table 1).

Beginning students may have difficulty figuring out what should be included in or omitted from a lab article. Instead of providing simple step-by-step, fill-in-the blank formats, we explain typical content of the various sections through regular in-class discussions of student papers that compare student writing to writing by practicing scientists. The instructor highlights what was done well, and how improvements can be made. Expectations increase throughout the course as the criteria for grades are adjusted to include the ideas discussed. In our experience, by the end of the first semester all the students are capable of describing their work in a way similar to a regular scientific paper but can reflect individual student's styles.

| Source | Accessibility |
|---|---|
| **EBSCOHost** | Subscribing public library or school |
| **Directory of Open Access Journals (doaj.org )** | Free |
| **Google Scholar (scholar.google.com )** | Free |
| **JSTOR (jstor.org )** | Free to search |
| **Pubmed (www.ncbi.nlm.nih.gov/pubmed )** | Free |
| **ScienceDirect (sciencedirect.com )** | Free to search |
| **arXiv.org** | Free |

**TABLE 1** Sample Databases

Peer review is an important learning activity. Students benefit as writers by having comments from additional readers and as scientists by seeing different experimental approaches and data analysis. Blinded and unblinded written peer review is also frequently used (the online platform Turnitin facilitates distribution of papers electronically for peer review through its PeerMark assignment feature). Students can provide critiques such as "insufficient introduction to the topic", "unclear motivation", "unclear writing", or "incorrect calculations". A portion of the lab grade is derived from a student's effort on peer review.

## 2.7 Harwood Model of Scientific Inquiry

We formulated the principles above by imagining a typical work flow of a new research student and have found that is compatible with Harwood model of inquiry (Harwood, 2004b,a) which is representative of how science is conducted as a series of iterative processes, and which is applicable across disciplines. Harwood specifies 10 activities found in a typical research process that we emphasize in our experiments: 1) Ask questions, 2) Define the problem, 3) Form the question, 4) Investigate the known, 5) Articulate an expectation, 6) Carry out the study, 7) Examine the results, 8) Reflect on the findings, 9) Communicate with others, and 10) Make observations.

# 3 INDEPENDENT RESEARCHER LABS FOR PHYSICS

The instructions provided allow students the freedom to develop their own approach to answer an experimental question. Methods of data analysis are discussed, but the final choice can vary from student to student. Labs are organized in modules over a three-week period (one 40-minute class period/week) according to the following schedule:

- Day 1: An experiment serves as an initial introduction to topic and corresponding instruments. Data analysis and/or additional reading assigned for homework
- Day 2: The main experiment. Data analysis and lab article preparation assigned for homework
- Day 3: Either a whole class lab article presentation by students followed by a discussion and comments on how to improve it. Alternatively, the third day can be used as time for experiments (either to complete experiment from Day 2, or for additional experiments if data analysis shows that a previous design does not produce good data)

Two examples of the honors physics laboratory modules (resistivity and Maxwell pendulum) that we have used in our classes are presented below. These two are typical labs for physics that are described in many sources (for example, in the AP Physics Lab Manual (The College Board, 2015), or on the Vernier Software Technology website (Vernier Software & Technology, 2017)), but have been redesigned for the goals outlined above. Here, the focus is "how" we present a task to students and what is expected from them.

In addition, a module for AP physics is presented based on a combination of experiments and computer simulations using Wolfram Mathematica. A sequence of seven 3-weeks lab modules (21 weeks total) with full code is described elsewhere (Samsonau, 2018).

## 3.1 Resistance and resistivity (Honors physics)

This lab is divided into two parts. In part one, students compare the nominal resistance of ohmic resistance to the experimental value and plot voltage (V) vs current (I) for both ohmic and non-ohmic (light bulb) resistors. They must apply at least 10 different voltages ranging from 0 to 15 V. The teacher provides limited instructions, such that students can progress, but have to think hard on their own how to design and perform an experiment. A teacher may assist students in assembling a circuit if needed. Students are initially provided with a rheostat (capable to handle current of 3-5 Amps, and having a maximum of 10-50 Ohms), a power supply with a short circuit protection (0-15V), mechanical ammeter (0-5A), mechanical voltmeter (0-15V), a knife switch, connecting wires, and a light bulb (rated to 1.5 V or similar). Students may ask for additional equipment.

In part two, students find the resistivity of an unknown material and give a hypothesis about the material used in the wire. Students are provided with a coil of thin wire of unknown material from which students can cut a piece; a mechanical voltmeter with ranges 3, 5, 15V; a mechanical ammeter (ranges of 1 and 5 A); a power supply of at least up to 15 V that can safely handle a short circuit; connecting wires; and a knife switch. In this part, students may ask for additional equipment, but are limited to the voltmeter and ammeter provided. Students must design the procedure and assemble the whole circuit by themselves. Important step for students is to recognize that 1) if they perform measurements on a short wire, they essentially create a short circuit, 2) a precision of instruments dictates the way an experiment shall be designed – for example they can choose a length of a wire which would allow them to take measurement of voltage with is considerably higher that an instrumental error.

In this example, the question and problem are provided by the instructor (as would be common in early stages of research), but students are responsible for the investigation and communicating results (Table 2)

### 3.2 Maxwell Pendulum (Honors physics)

In this lab students are asked to study energy transfer in Maxwell wheel pendulum. They must develop their own experiment, design and build a device, and then perform experiments.

There are no specific instructions given to students to set up the question or evaluate the result. Typically, this lab occurs at the end of the first semester when students have significant experience in addressing such open-ended tasks.

Students have access to all resources in the physics and engineering labs. Some students chose to use such equipment as sparkling timer, photogate, ultrasonic motion sensor, or slow-motion video camera on smartphone. They must account for the properties and limitations of these sensors while designing their pendulums. For example: to use a photogate, students have to make slits in the pendulum; to use an ultrasonic sensor, the pendulum must not be very thin in order for the sensor to pick it up.

### 3.3 Motion with a drag (AP Physics)

The teacher introduces students to Mathematica and covers basic operations, plotting, and solving equations. In class, students are asked to describe free fall with the kinematic equation and to plot it. As a homework assignment, students learn how to use NDSolve to solve differential equations, and read general theory related to Euler and Runge-Kutta methods.

The primary objective in this experiment is to study motion with a drag with several potential subtasks:

- Model free fall motion with a drag
- Determine which drag type to use (turbulent vs laminar flow)
- Make corresponding plots
- Design and perform an experiment to evaluate validity of the numerical simulation

On the first day a teacher gives a basic example of use of NDSolve function with free fall in no-air-resistance situation, and this solution is compared first to the kinematic equation, then to experimental results by dropping a heavy ball. As a first part of a homework, students will make a Mathematica program to integrate an equation of motion with a drag. They should (read and) decide which drag to use: laminar or turbulent. For the second part of the homework, students have to design an experiment to measure an effect of a drag. For example, one can use water in a high and wide graduated cylinder, and either drop a small object to sink, or produce bubbles to observe how they rise.

On the second class of the module, students proceed to experimental work. For homework, students analyze their experimental data and use Mathematica to model the parameters of their actual experiment to compare results. They consider if the designed and completed experiment allows the effects of drag to be observed. After they compare the results of the numerical simulation with the real experiment, they may need to redesign experiment and redo it, what can be done on the third day of the module.

| Harwood Task | Resistance and Resistivity | Maxwell Pendulum | Motion with a drag |
|---|---|---|---|
| Ask question (general) | *By teacher:* Measuring an unknown property of a material. | *By teacher:* Evaluate validity of law of conservation of energy. | *By teacher:* Build computer simulation and evaluate how close it describes real phenomenon. |
| Define the problem (focusing exploration) | *By teacher:* determine resistivity of wire of unknown material. | *By teacher:* Do this using self-constructed Maxwell pendulum. | Students choose specific conditions, object and media for experiments. |
| Form the question (question drives study) | *By teacher:* Test resistance and geometry of wire using provided equipment. Calculate resistivity and evaluate uncertainty. | Implement one of many possible designs of Maxwell pendulum (depends on how students choose to measure position and speed. | Students specify values to measure and compare to simulations. They set up a procedure of comparison. |
| Investigate the known | To estimate expected value, students identify typical materials used in wires and resistivity. | Students learn about various ways to do Maxwell pendulum and types of sensors available. | Students learn how computer simulations are performed, as well as how drag acts in different conditions. |
| Articulate an expectation | Students shall clearly state the excepted range for resistivity and explain why. | Students can evaluate precision of the sensors and evaluate potential precision of final experiment. | Students form quantitative expectation for behavior of a real system from computer simulation. |
| Carry out the study | Design and perform experiment. Collect data. | Design and make device. Design and perform experiment and analyze data. | Students design and perform computer simulations, experiment, and data analysis. |
| Examine the results | Does collected data look meaningful? Is error small enough? Should experiment be redesigned and repeated? | | |
| Reflect on the findings | What material do you think the wire is made of? | According to your experiment, can you reject the law of energy conservation? | How well simulations describe the experimental observations? |
| Communicate with others | Prepare lab article, do peer-review, do class presentation. | | Prepare lab article, do peer-review. |
| Make observations | During the work students have to notice "unusual" behavior of equipment and realize that it was due to very low resistance of a short wire (students typically start with a short wire). | What can you do to reduce error in your experiment? What was the largest source of error? Students can be encouraged to report any other interesting observations. | It is highly likely that a first design of an experiment will not deliver good enough data. Students have to notice what should be improved and do that. |

**TABLE 2** Summary of Physics Labs

## 4 | INDEPENDENT RESEARCHER LABS FOR CHEMISTRY

Chemical educators such as Miles Pickering at Princeton University (Kandel, 1997; Pickering, 1985) and James Spencer at Franklin & Marshall College (Farrell et al., 1999; Spencer, 2006) were early proponents for inquiry and student directed learning at the undergraduate level. The lessons and instructional material have filtered down to the secondary school level, though these are often constrained by high school laboratory facilities and periods, cost, and teacher expertise.

Examples that we have successfully implemented in our chemistry curriculum encompass typical chemical skills such as synthesis and quantitative analysis. The focus is on teaching standard experimental techniques as well as introducing students to the realistic uncertainty, imprecision, and failure that accompanies open inquiry (Firestein, 2015). In that sense, some students may have results that appear to be "less successful" than their peers; the challenge is to get them to accept the data that result from their experiment (Taylor, 2017). Data from all lab groups is available to students (typically via Google Sheets and Google Docs) to facilitate discussions about variability and drawing conclusions.

It is not imperative that every lab incorporate all aspects of open-inquiry, and the teacher may provide the question and general methods for many experiments geared towards beginning students. However, the experiments are designed so students experience aspects that are common in authentic research and described by the Harwood methodology as summarized in Table 3.

| Harwood Task | Heat of Combustion | Analysis of Biodiesel | Vitamin C Analysis |
|---|---|---|---|
| Ask questions | | | |
| Define the problem | General question provided by the teacher. | | |
| Form the question | | | |
| Investigate the known | Students review literature and determine that no consensus value is known. | Typical values for biodiesel range from 32 to 45 kJ/g. | Target range provided by bottle. |
| Articulate an expectation | Students expect values of about 40 kJ/g. | Energy content will vary based on proportion of saturated and unsaturated fats. | Students expect values between 970 mg to 1030 mg. |
| Carry out the study | General calorimetry equipment is provided; students can modify to prevent heat loss. | | Students choose analytical method for titration. |
| Examine the results | Equipment intentionally limits precision and accuracy. Students consider if experiment can be re-designed to improve data. | | Students evaluate data to determine if additional trials are necessary or if alternative method should be used. |
| Reflect on the findings | Qualitative demonstration of conservation of energy. | Values differ for different types of oils. | Out-of-range values may be due to analytical method. |
| Communicate with others | Students must discuss results to obtain enough data points for meaningful analysis and to prepare written report for peer review. | | Students summarize results in a memo to head of analysis lab. |
| Make observations | Different color candles had different values. Values are smaller than anticipated. | Energy content varies for oil type. | Tablets from single batch have variability; method of analysis can affect results. |

**TABLE 3** Summary of Chemistry Labs

## 4.1 | Heat of Combustion

In a common introductory lab, students measure the heat of combustion of a candle (Backus, 2005; Inv, 2012); while not specific, most procedures suggest that the candle is paraffin wax comprised primarily of long-chain alkanes with the general formula $C_nH_{2n+2}$ (n =19–36).

Despite the ubiquitousness of paraffin wax, there is surprisingly little literature data for the heat of combustion. The primary citation reports the average heat of combustion to be about 43 kJ/g (Hamins, 2005), though the value commonly given in many lab manual instructions is 42 kJ/g. To further complicate matters, differing values (41.57 kJ/g and 47.86 kJ/g respectively) have been reported for a yellow votive candle and a white emergency candle (Rettich et al., 1988). Of course, the value for any individual candle will depend on its precise hydrocarbon makeup, and candles could be alternatively composed of beeswax, or soy-based wax (Cottom, 2000; Hoffmann et al., 2014).

This uncertainty is a benefit in our system. A seemingly simple experiment can be "complicated" by limiting students to equipment and materials provided; in our case, they only had a ring stand, iron ring, and two different sized aluminum cans. When students measured the heat of combustion of typical colored 3-inch birthday candles using basic calorimetric techniques, they obtained values between 20-25 kJ/g—the correct order of magnitude though only about 50% of the actual heat of combustion. There were subtle differences observed between different colors of candles as well. Since each lab group had data for only a single candle, students pooled data to begin a discussion of the law of conservation of energy—if energy is neither lost not created, the transfer between the candle flame and water must be inefficient. Students can develop procedures to reduce the systematic error in the experimental apparatus as well as to determine if the apparent differences between candle colors is due to the presence of an additive or just an artifact of intergroup variability.

**4.2 |Synthesis and Analysis of Biodiesel**

The preparation of biodiesel from organic matter has become a popular activity in many classrooms (Clarke et al., 2006; De La Rosa et al., 2014; Yang et al., 2013), and is well suited to our methods. The general preparation requires heating a sample of oil with methanol and aqueous sodium hydroxide, and is readily adapted to different plant oils, including canola, corn, olive, soybean, and coconut oils. These oils differ in their composition of saturated and unsaturated fatty acids (Table 4).

| Plant Oil | % Saturated Fatty Acids | % Unsaturated Fatty Acids | |
|---|---|---|---|
| | | Monounsaturated | linoleic and linolenic |
| coconut | 76.2 | 7.5 | 0 |
| corn | 14.6 | 49.6 | 34.3 |
| soybean | 12.5 | 28.9 | 55.9 |
| olive | 9.2 | 84.4 | 4.6 |
| Data from (Bettelheim et al., 2016) | | | |

**TABLE 4** Average Fatty Acid Composition of Plant Oils

Energy contents of various biodiesel fuels obtained under ideal conditions are generally between 32 to 45 kJ/g, with fuels containing amounts of saturated fatty acids typically having higher energy content (Ozcanli et al., 2013). Students using a modification of a standard procedure (Tripp and McKenzie, 2015) obtained values that were expectedly low for at least two reasons: no attempt was made to quantify or maximize the synthetic yield and the process for measuring energy content was subject to inefficient heat transfer. However, the raw data produced by the students provided the basis to discuss experimental technique and significant figures (Table 5). One group chose to make duplicate measurements and obtained significantly different values for coconut oil biodiesel; on the other hand, two different groups arrived at similar results for soybean oil biodiesel.

| Biodiesel Source | mass water (g) | mass biodiesel (g) | Δ T (K) | energy content (kJ/g) | |
|---|---|---|---|---|---|
| | | | | Experimental | Literature |
| Coconut oil *a* | 154.98 | 1.1 | 20 | 11.79 | 35.00 *c* |
| | 191.13 | 2.02 | 20 | 7.92 | |
| Soybean oil[b] | 65.90 | 0.47 | 20.30 | 11.91 | 37.30 – 37.68 *c* |
| | 76.2 | 0.53 | 20 | 12.03 | |
| Olive Oil | 74.7 | 0.23 | 20 | 27.3 | 32.781 *d* |

Precision and significant figures are reproduced verbatim from student data. a. Duplicate values obtained by single lab group. b. Values obtained by two separate lab groups. c. Data from Mehta and Anand (2009). d. Data from Ozcanli et al. (2013).

**TABLE 5** Representative Raw Data

### 4.3 Vitamin C Analysis

Many chemistry laboratory manuals offer procedures to analyze or quantify components of consumer products such as hydrogen peroxide or analgesics (The College Board, 2014). Such experiments fit within our framework as they provide students the opportunity to do realistic chemistry tasks as well as obtain data that may not have a clear "right answer." For example, though many over-the-counter medications and vitamins state the amount of active ingredient, any individual tablet may have between 97 to 103% of the stated label claim. In addition, any products past the expiry date may have less due to potential decomposition. Students were provided with a general procedure to measure the amount of ascorbic acid in a sample using a starch-iodine titration (Tripp and McKenzie, 2015) and given the following scenario:

> *Pretend that you are a chemist analyzing a company's products: a vitamin C pill that should contain 1000 mg of ascorbic acid (three separate pills will be tested to determine how consistent the factory is manufacturing the tablets) and pineapple juice that claims to have 100% of the daily recommendation for vitamin C (for adults, this is 75 to 100 mg).*

Some students chose to use standard burets while others used a simplistic drop quantification and counting method. Each lab group analyzed a separate tablet in triplicate and then results were shared.

Notably, the drop-counting method (tablet #2) gave similar results to the burette method (tablet #1) in one case, while significantly different results in the other case (Table 6).

| Sample | Amount of Vitamin C (g) | Difference from claim |
|---|---|---|
| Tablet #1 (titration) | 1.018 | 1.82% |
| Tablet #2 (drop-counting) | 1.019 | 1.88% |
| Tablet #3 (drop-counting) | 1.312 | 31.20% |
| Data from three separate student groups. The drop-counting method is known to be highly variable (Ealy and Pickering, 1991). | | |

**TABLE 6** Student Data for Vitamin C Analysis

## 5 | CONCLUSION

We expect our students to use the experience with working on tasks with no predetermined results to learn the difference between ideal values specified by manufacturer and actual values, and to become comfortable working with the "unknown". By allowing students responsibility for designing experimental procedures, a student realizes that research is not merely following a well-worn path and that the individual must decide when an experiment is good enough rather than if they got the "right answer". Students see that experiments and science research in general is an iterative process, learn to evaluate sources of errors, and identify steps to improve an experiment. Finally, students are expected to make mistakes and experience frustration, and overcome these. This is an important lesson not only for scientific research, but also for life in general.

With the number of students who present their work at research competitions, such as the Intel International Science and Engineering Fair, Regeneron Science Talent Search, and the International and USA Young Physicist Tournaments, we feel that our approach is useful not only for schools who have implemented their own research program, but also for schools aiming to ignite and support students' interest in authentic research activity.